\documentclass[onecolumn,10pt,a4paper,sort&compress,aps,pra,showpacs,superscriptaddress]{revtex4-2}

\usepackage[english]{babel}
\usepackage[caption=false]{subfig}
\addto\captionsenglish{\renewcommand{\figurename}{FIG.}}

\usepackage{amsmath,amssymb,amsthm}
\usepackage{graphicx}
\usepackage{dcolumn}
\usepackage{bm}
\usepackage{hyperref}
\usepackage{siunitx}
\DeclareSIUnit\Molar{\textsc{M}}
\usepackage[capitalise,nameinlink]{cleveref}
\usepackage{natbib}

\usepackage[normalem]{ulem}
\usepackage{comment}
\usepackage[usenames,dvipsnames]{color}
\usepackage[mathlines,displaymath,switch]{lineno}

\newcommand*\patchAmsMathEnvironmentForLineno[1]{%
  \expandafter\let\csname old#1\expandafter\endcsname\csname #1\endcsname
  \expandafter\let\csname oldend#1\expandafter\endcsname\csname end#1\endcsname
  \renewenvironment{#1}%
     {\linenomath\csname old#1\endcsname}%
     {\csname oldend#1\endcsname\endlinenomath}}%
\newcommand*\patchBothAmsMathEnvironmentsForLineno[1]{%
  \patchAmsMathEnvironmentForLineno{#1}%
  \patchAmsMathEnvironmentForLineno{#1*}}%
\AtBeginDocument{%
\patchBothAmsMathEnvironmentsForLineno{equation}%
\patchBothAmsMathEnvironmentsForLineno{align}%
\patchBothAmsMathEnvironmentsForLineno{flalign}%
\patchBothAmsMathEnvironmentsForLineno{alignat}%
\patchBothAmsMathEnvironmentsForLineno{gather}%
\patchBothAmsMathEnvironmentsForLineno{multline}%
}

\newcommand{\changed}[1]{{\color{blue}#1}}
\renewcommand{\changed}[1]{#1}

\renewcommand{\vec}[1]{\bm{#1}}
\newcommand{\intd}{\mathop{}\!\mathrm{d}}
\newcommand{\abs}[1]{\left\lvert{#1}\right\rvert}

\newcommand{\Rey}{\operatorname{\mathit{R\kern-.04em e}}}
\newcommand{\dOmega}{\partial\Omega}

\newcommand{\dt}{\mathrm{d}t}

\newcommand{\headtail}[1][i]{\vec{\hat{x}}_{0}^{#1}}
\newcommand{\heading}{\theta}
\newcommand{\sep}{h}
\newcommand{\eij}[1][ij]{\vec{\hat{e}}_{x_{#1}}}
\newcommand{\htheta}{$\sep{}$--$\heading{}$}

\begin{document}
	
\title{The pairwise hydrodynamic interactions of synchronized spermatozoa}

\author{Benjamin J. Walker}
\email{Corresponding author: benjamin.walker@maths.ox.ac.uk}
\affiliation{Wolfson Centre for Mathematical Biology, Mathematical Institute, University of Oxford, Oxford, OX2 6GG, UK}

\author{Kenta Ishimoto}
\email{ishimoto@ms.u-tokyo.ac.jp}
\affiliation{Graduate School of Mathematical Sciences, The University of Tokyo, Tokyo, 153-8914, Japan}

\author{Eamonn A. Gaffney}
\email{gaffney@maths.ox.ac.uk}
\affiliation{Wolfson Centre for Mathematical Biology, Mathematical Institute, University of Oxford, Oxford, OX2 6GG, UK}


\date{\today}

\begin{abstract}
The journey of mammalian spermatozoa in nature is well-known to be reliant on
their individual motility. Often swimming in crowded microenvironments, the
progress of any single swimmer is likely dependent on their interactions with
other nearby swimmers. Whilst the complex dynamics of lone spermatozoa have
been well-studied, the detailed effects of hydrodynamic interactions between
neighbors remain unclear, with inherent nonlinearity in the \changed{pairwise
dynamics} and potential dependence on the details of swimmer morphology. In
this study we will attempt to elucidate the pairwise swimming behaviors of
virtual spermatozoa, forming a computational representation of an unbound
swimming pair and evaluating the details of their interactions via a
high-accuracy boundary element method. We have explored extensive regions of
parameter space to determine the pairwise interactions of synchronized
spermatozoa, with synchronized swimmers often being noted in experimental
observations, and have found that two-dimensional reduced autonomous dynamical
systems capture the anisotropic nature of the swimming speed and stability
arising from near-field hydrodynamic interactions. Focusing on two
\changed{initial} configurations of spermatozoa, namely those with swimmers
located side-by-side or above and below one another, we have found that
side-by-side cells attract each other, and the trajectories in the phase plane
are well captured by a recently-proposed coarse-graining method of
microswimmer dynamics via superposed \changed{regularized} Stokeslets. In
contrast, the above-below pair exhibit a remarkable stable pairwise swimming
behavior, corresponding to a stable configuration of the plane autonomous
system with swimmers lying approximately parallel to one another. At further
reduced swimmer separations we additionally observe a marked increase in
swimming velocity over individual swimmers in the bulk, potentially suggesting
a competitive advantage to cooperative swimming. These latter observations are
not captured by the coarse-grained \changed{regularized} Stokeslet modeling
\changed{or simple singularity representations}, emphasizing the complexity of
near-field cell-cell hydrodynamic interactions and their inherent anisotropy.
\end{abstract}

\maketitle

\section{Introduction}
\label{sec:intro}
Rarely found in isolation, mammalian spermatozoa commonly swim as part of
large cell populations in a variety of artificial and biological
microenvironments \cite{Gray1928,Rothschild1949,Moore2002,Creppy2015}. Indeed,
a single human spermatozoon can be one of hundreds of millions of individuals
in a single fertile ejaculate \cite{Zinaman2000}. Despite the relevance of
such circumstances to areas of scientific research such as fertility
diagnostics, where an understanding of the collective behaviors of
spermatozoa may be applicable to assisted reproductive therapies and enable
more-efficient experimentation \cite{Lai2012,Swain2013,Tung2014}, many
investigations have considered only the behaviors of a lone spermatozoon.
Such studies range from examining the motion of individual swimmers in
non-trivial background flows to their behavior in confined geometries
\cite{Ishimoto2015,Ishimoto2014,Friedrich2010,Elgeti2011,Fauci1995,Smith2009}.
However, there have been comparatively few theoretical studies that have
considered the collective behaviors of multiple spermatozoa, recent examples
being the large-scale numerical simulations of \citet{Schoeller2018,Yang2010},
with notable additional work pertaining to swimming in two dimensions, though
there remains significant scope for investigating the complex near-field
interactions of swimmers in detail. The latter investigations include
classical two-dimensional swimming sheet models, introduced in the classical
study of \citet{Taylor1951} and used later by \citet{Fauci1995} to study the
interactions of model spermatozoa via the immersed boundary method with
two-dimensional hydrodynamics. Restricted to motion in a plane, but capturing
three-dimensional hydrodynamics, \citet{Yang2008} simulated swimmer
interactions via multiparticle collision dynamics, utilizing a simple bead
model of spermatozoa. A key result of their study was the observation of
hydrodynamic attraction between two neighboring spermatozoa, affirmed by the
computations of \citet{Simons2015,Llopis2013}. These studies allowed for
swimmer motion in three dimensions, and were based on \changed{regularized} Stokeslet
slender body theory and bead and spring models respectively, neglecting
potential hydrodynamic effects of the swimmer cell body that may not be
captured by these methodologies. Interactions between swimmers and boundaries
have previously been reported to be highly sensitive to swimmer morphology
\cite{Ishimoto2013,Walker2019}, thus it is not clear a priori that the
pairwise interactions of spermatozoa will be independent of their geometry.
Hence as a first objective of this study we will aim to elucidate any
behaviors of attraction and repulsion of neighboring spermatozoa which are
not bound without confining the cell swimming dynamics to a plane, utilizing a
high-accuracy boundary element method to capture the swimmer cell body.

Studying swimmers in configurations directly above or beside one another and
considering elastohydrodynamics via a bead-and-spring model,
\citet{Llopis2013} found that pairwise swimming was unstable for their model
swimmers. Further to remarking on the attraction and repulsion of their driven
bead chains, \citeauthor{Llopis2013} note a surprising dependence of swimming
speed on swimmer configuration, with swimming speed increasing only for
swimmers directly above one another, and decreasing for side-by-side swimmers.
The experimental study of \citet{Woolley2009} concerning bovine spermatozoa
supports this conclusion that relative swimmer position impacts on changes to
the speed of progression effected by proximity, though detailed theoretical
exploration of this result accounting for complex hydrodynamics and
spermatozoan geometry is, to the best of our knowledge, lacking. Thus as a
second objective of this study we will aim to determine the effects of swimmer
configuration on swimming speed from a kinematic representation of
spermatozoa, in particular noting if the inclusion of the non-trivial geometry
of human spermatozoa invalidates qualitative conclusions drawn from simpler
models.

Also significant to the interaction of nearby spermatozoa is the phenomenon of
flagellar synchronization, in which the coupling of hydrodynamics and filament
mechanics can result in out-of-phase beating swimmers rapidly adopting a
common beat pattern and phase. Synchronization has been observed in various
flagellated organisms \cite{Brumley2014,Woolley2009,Friedrich2014}, with
extensive theoretical study supporting the role of hydrodynamics in filament
synchronization
\cite{Yang2008,Uchida2017,Elfring2011,Goldstein2016,Fauci1990,Olson2015}.
However, in this study we will consider spermatozoa swimming in very close
proximity, with separations typically being less than one cell length, and
hence we will assume throughout that flagellar beating has been synchronized.
In turn, this circumvents the need for elastohydrodynamic calculations, which
generally suffer from severe numerical stiffness, preventing detailed
exploration of the parameter space associated with the numerous degrees of
freedom characterizing the relative motion of two interacting sperm cells,
whilst exploring extensive regions of parameter space underpins the aims of
this work.

There has been significant study pertaining to the collective behaviors of
swimmers, ranging from the characteristics of bacterial suspensions to active
colloidal particles
\cite{Sokolov2012,Zottl2013,Saintillan2007,Oyama2016,Li2016,Ezhilan2013,Marchetti2013}.
The flow field surrounding an individual spermatozoon is often approximated by
that of the \changed{force dipole}, with constant singularity strength of a
pusher-type swimmer \cite{Lauga2009}. The pairwise interactions between such
simple model swimmers results in hydrodynamic attraction irrespective of the
configuration of the swimmers, owing to the symmetry of the flow field
representation, though this may not be realistic due to the noted dependence
of swimming on cell morphology \cite{Pooley2007}. Recently a theoretical
coarse-grained approach for the simulation of swimmer populations was
suggested by \citet{Ishimoto2017b}, and later implemented by
\citet{Ishimoto2018}, seeking to refine the most-basic singularity
representation of a spermatozoon. This methodology aims to approximate the
flow field induced by a lone swimmer by a small number of \changed{regularized}
singularities \cite{Cortez2001} with time-varying coefficients for use in
large-scale simulations, seeking to improve upon simple far-field singularity
representations. \citeauthor{Ishimoto2018} simulated a population of
spermatozoa moving in two dimensions, with the coarse-grained model
well-distinguishing between spermatozoan clustering behaviors emergent from
the prescription of different flagellar waveforms, with detailed waveforms
representing those that may be observed in spermatozoa in various rheological
media. Further, a time-averaged representation of the swimmers fails to
reproduce such qualitative differences in the population-level dynamics,
highlighting the significance of temporally-fluctuating hydrodynamic
interactions in the collective behavior of spermatozoa. The efficacy of
utilizing their approach in order to qualitatively capture short range
swimmer-swimmer interactions is unknown, as is the ability of the methodology
to approximate dynamics in three spatial dimensions. Therefore, as an
additional objective of this study we will evaluate the success of this
approach when it is applied to three-dimensional motion, comparing against
boundary element computations and ascertaining the degree of qualitative
agreement between the two models. As a final objective we will aim to
interpret the computational results of the coarse-grained model in terms of
fundamental singularity solutions of Stokes equations, assessing the utility
and accuracy of such simple descriptions in approximating swimmer-swimmer
interactions.

Hence, in this paper we will examine in detail the hydrodynamic interactions
of a pair of virtual spermatozoa, represented by an idealized but
representative and commonly-used geometry. Prescribing synchronized flagellar
kinematics, we will simulate the motion of our virtual swimmers using a
high-accuracy boundary element method, aiming to quantify pairwise swimming
behaviors in detail, with particular focus on the direction and speeds of
relative progression of the pair. After demonstrating that relative swimmer
motion may be reduced to an approximate two-dimensional dynamical system, we
analyze such a system to discern long-time swimming behaviors. Further, we
will draw comparison of simulated behaviors with established experimental
results, and determine the level of accuracy to which the simple swimmer
representations of \citet{Ishimoto2018} predict the interactions of virtual
spermatozoa.
\section{Methods}
\label{sec:methods}
\subsection{Virtual spermatozoa}
\label{sec:methods:virtual_sperm}
In order to model the interactions of two spermatozoa we introduce the
neutrally-buoyant \emph{virtual kinematic spermatozoon} following
\citet{Ishimoto2014}, formed of an ellipsoidal-like head connected to a long
slender flagellum. For the head we adopt the idealized geometry of
\citet{Smith2009}, forming a computational triangular mesh of the surface as
parameterized in the \cref{app:head_parameterisation}. This is given with
respect to a swimmer-fixed frame with right-handed orthogonal coordinate axes
$x_{i1}x_{i2}x_{i3}$, where $i\in\{1,2\}$ labels each individual swimmer.
These axes are taken such that the $x_{i1}$$x_{i2}$ plane coincides with the
plane of the flagellar beat of swimmer $i$, with the corresponding head-tail
junction situated at $(x_{i1},x_{i2},x_{i3})=\vec{0}$, where the head-tail
junctions have locations $\headtail{}$ in the inertial laboratory frame
$\hat{x}_1\hat{x}_2\hat{x}_3$. Unit vectors of the swimmer-fixed frames
expressed in the laboratory frame are denoted by $\eij$ for swimmer
$i\in\{1,2\}$ and direction $j\in\{1,2,3\}$.

Each flagellum is represented by a spherically-capped cylinder of radius
0.125\si{\um} and length $L_{\text{f}}=56\si{\um}$, \changed{a standard scale
for human spermatozoa \cite{Cummins1985},} and their motion is prescribed in their respective
swimmer-fixed frames. Here we utilize the representative beat pattern of
\citet{Dresdner1981} as adopted in previous studies of spermatozoon behavior
\cite{Ishimoto2014}, with the flagellar beat parameterized by
$s\in[0,s^{\star}]$ as
\begin{align*}\label{eq:methods:flag_beat}
x_{i1}' &= s\,, \\
x_{i2}' &= (A - B s) \sin(ks - \omega t + \psi_0(i)) \,,\\
x_{i3}' &= 0\,,
\end{align*}
for wavenumber $k$, frequency $\omega$, and phase shift $\psi_0(i)$, where we
take $(A,B) = (0.0543L_{\text{f}},0.1087)$ for flagellum length
$L_{\text{f}}=56\si{\um}$. Here $x_{i1}'x_{i2}'x_{i3}'$ denotes a
flagellum-fixed reference frame \cite{Dresdner1981,Smith2009}, related to the
respective swimmer-fixed frame by a time-dependent rotation and translation
that ensures that the local flagellum tangent at the base of the flagellum is
parallel to $\eij[i1]$, consistent with previous computational studies of
human spermatozoa \cite{Ishimoto2014,Smith2009}. The quantity $s^{\star}$ is
taken such that the total length of the flagellum is constant, whilst the
angular frequency of the beating is fixed such that $\omega/2\pi=14\si{\Hz}$,
in alignment with observations of spermatozoa in Newtonian media
\cite{Smith2009a}. We note that the results that follow are qualitatively
independent of this chosen beat frequency, as adjustments to this frequency
simply correspond to performing a rescaling in time. Further, throughout we
will consider flagellar beating with wavenumber $k=3\pi/L_\text{f}$ as in
\citet{Dresdner1981}, though we remark that the presented results are not
qualitatively affected by small changes in wavenumber. With reference to the
widely-documented synchronization of flagellar beating when swimmers are in
close proximity, as remarked in \cref{sec:intro}, we prescribe $\psi_0(1) =
\psi_0(2) = \psi_0$ for some constant phase shift $\psi_0\in[0,2\pi)$, and
hereafter assume in this kinematic study that the beating of nearby swimmers
is synchronous. For convenience we take $\psi_0$ such that the free-space
headings of the swimmers are parallel to $\eij[11]$ and $\eij[21]$
respectively at time $t=0$. Here and throughout the term \emph{free-space
heading} refers to the direction of net progression over a beat cycle for a
lone swimmer in free-space, and is used as a reference from which we describe
the orientation of a swimmer.

\begin{figure}[t]
  \centering
  \subfloat[\label{fig:methods:mesh:head}]{%
    \includegraphics[width = 0.4\textwidth]{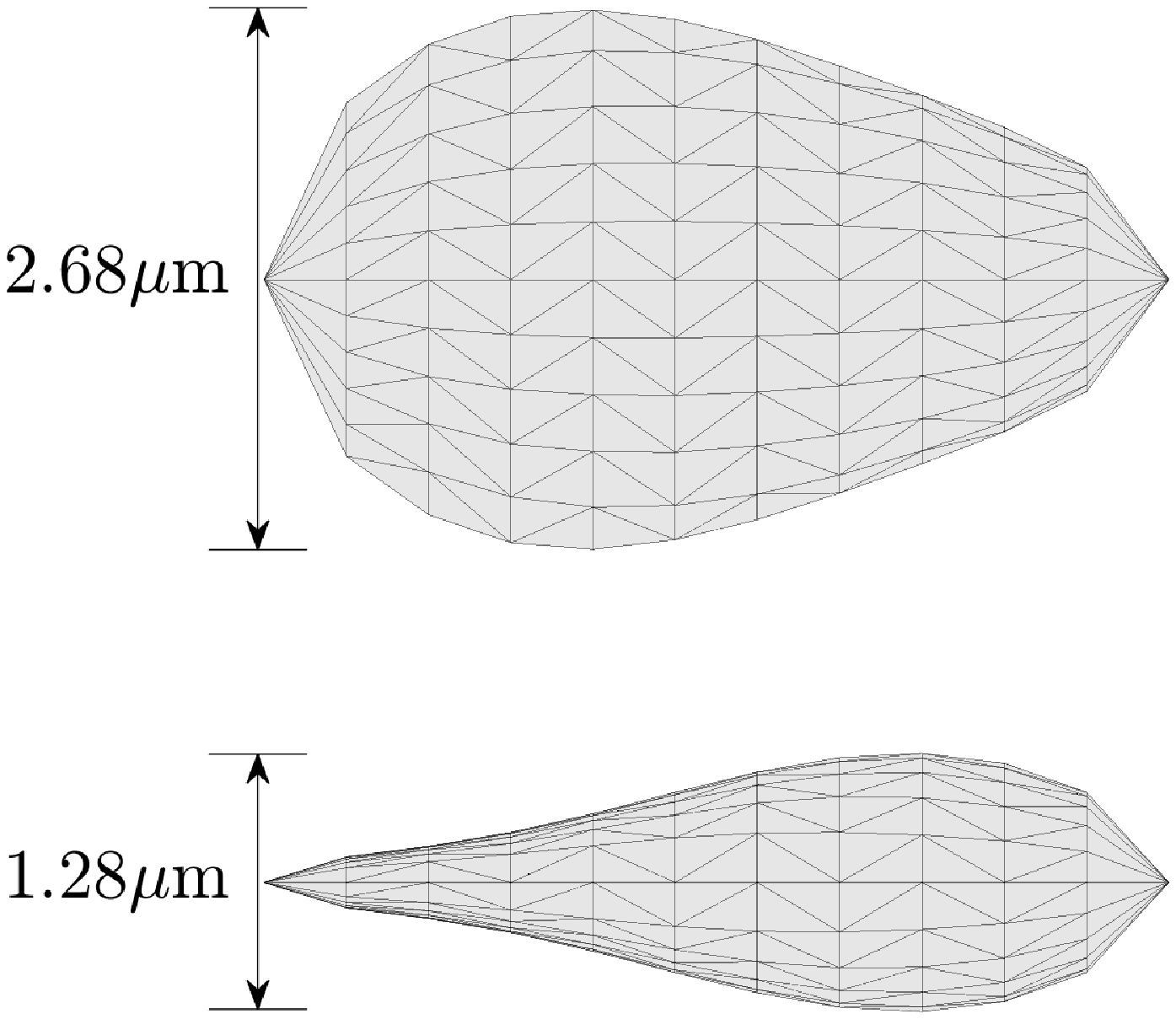}}
  \subfloat[\label{fig:methods:mesh:adaptive}]{%
    \raisebox{1.5em}{\includegraphics[width = 0.4\textwidth]{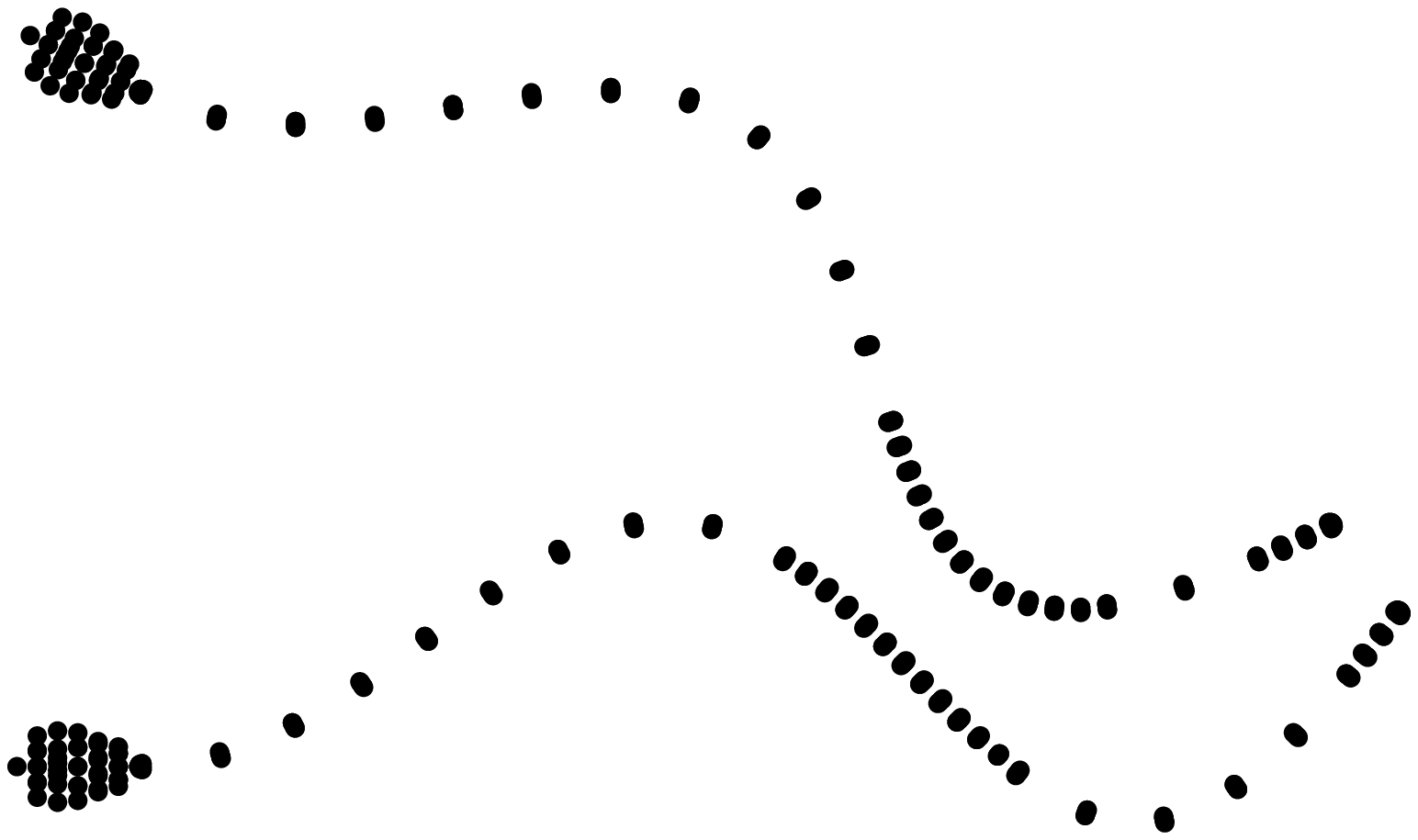}}}
  \caption{\label{fig:methods:mesh} Computational meshes representing
  spermatozoan geometry. \protect\subref{fig:methods:mesh:head} Views of a
  triangular mesh representing the head of a spermatozoon, with vertices
  created in ellipsoidal bands and equally spaced around each ellipsoidal
  cross section, following the parameterization given in the
  \cref{app:head_parameterisation}.
  \protect\subref{fig:methods:mesh:adaptive} An example of adaptive meshing
  based on swimmer proximity, with vertices shown as points and elements
  omitted. The distal portions of the flagella of both swimmers have been
  partially re-meshed to increase local accuracy whilst retaining
  computational efficiency. The mesh is shown at reduced resolution for visual
  clarity, though the numerous vertices at each cross section of the flagellum
  are effectively amalgamated into one ellipsoidal marker point due to the
  resolution of the image. \changed{Finally, note that these flagella
  correspond to a flagellum length of 56\si{\um}, a standard scale for human
  spermatozoa \cite{Cummins1985}.}}
\end{figure}

\subsection{Governing equations and numerical solution}
\label{sec:methods:governing_equations}
The microscale swimming of spermatozoa in an incompressible Newtonian fluid is
governed by the three-dimensional Stokes equations, which for pressure $p$ and
fluid velocity $\vec{u}$, expressed in the laboratory frame, are given by
\begin{equation}
  \mu\nabla^2\vec{u} = \nabla p\,, \quad \nabla\cdot\vec{u} = 0\,,
\end{equation}
where $\mu$ is the viscosity of the fluid and these equations hold throughout
the fluid domain $\Omega$ with boundary $\dOmega{}$. Here we take the fluid
domain to be the exterior of the virtual spermatozoa, and apply a no-slip
boundary condition on the swimmer surface, additionally enforcing that the
fluid velocity decays to zero at spatial infinity in the absence of any
non-trivial background flow. Following the boundary element formulation of
\citet{pozrikidis1992,pozrikidis2002}, the solution for the fluid velocity
$\vec{u}$ relative to the laboratory frame at a point $x^{\star}$ on the
boundary $\dOmega{}$ is given by
\begin{equation}
  u_j(\vec{x}^{\star}) = - \frac{1}{4\pi\mu}\int\limits_{\dOmega{}} G_{ij}(\vec{x},\vec{x}^{\star})f_i(\vec{x}) \intd{S}(\vec{x}) 
  + \frac{1}{4\pi}\int\limits_{\dOmega{}}^{PV} u_i(\vec{x})T_{ijk}(\vec{x},\vec{x}^{\star})n_k(\vec{x})\intd{S}(\vec{x})\,,
\end{equation}
where $i,j,k\in\{1,2,3\}$ and Einstein summation convention is assumed. Here
$G_{ij}$ and $T_{ijk}$ are the standard free-space velocity and stress Green's
functions of Stokes flow, which inherently satisfy the condition of far-field
decay, $\vec{n}$ is the surface normal directed into the fluid, $\vec{f}$
denotes the surface traction, and $\int^{PV}$ denotes a Cauchy principal value
integral. Application of the no-slip condition at the nodes of the
computational mesh representing the surface of the swimmers yields a linear
system of equations in the a priori unknown surface tractions $\vec{f}$ and
the cell linear and angular velocities, where the angular velocity of each
cell is measured relative to its own head-tail junction. Closure of this
underspecified system results from imposing force and torque--free conditions
on each swimmer, in addition to removing the gauge freedom in pressure by
enforcing that the mean of the normal component of the traction over the
domain boundary be zero, without loss of generality.

Given an initial swimmer configuration and prescribed flagellar kinematics, the
instantaneous linear and angular velocities of the swimmers are computed and
swimmer positions and orientations updated via a predictor-corrector scheme,
Heun's method, with timestep $\dt{}$, which for swimmer position
$\vec{\hat{x}}_0^{i}(t)$ and linear velocity $\vec{U}^i(t)$ is given explicitly by
\begin{equation}
  \vec{\hat{x}}_0^{i}(t+\dt{}) = \vec{\hat{x}}_0^{i}(t) + \frac{\dt{}}{2}\left[\vec{U}^{i}(t)
  + \vec{U}^{i}(t+\dt{})\right]\,.
\end{equation}
In this study we focus on the hydrodynamic interactions of the swimming pair,
and do not include in our model any swimmer-swimmer contact interactions,
halting numerical simulations prior to collision, noting that both short range
attractive and repulsive forces are of physical relevance but are typically
only modeled phenomenologically
\cite{Klein2003,Ishimoto2016a,Walker2019,Simons2015}. Computational meshes
consisting of $2,848$ triangular elements per swimmer are adaptively refined
based on swimmer proximity, with quantities linearly interpolated over
elements. Example meshes demonstrating swimmer head geometry and
proximity-based adaptive meshing are shown in \cref{fig:methods:mesh}. The
suitability of computational meshes was established by numerical convergence,
with numerical procedures being validated against the boundary element methods
of our previous studies \cite{Walker2019,Ishimoto2014}. The timestep $\dt{}$
is typically taken such that there are 100 timesteps per beating period,
though temporal refinement is increased by an order of magnitude when swimmers
are in very close proximity.

\subsection{Swimmer configurations}
\label{sec:methods:configurations}
\begin{figure}[t]
  \centering
  \subfloat[\label{fig:methods:configurations:side-by-side}]{%
    \includegraphics[width = 0.4\textwidth]{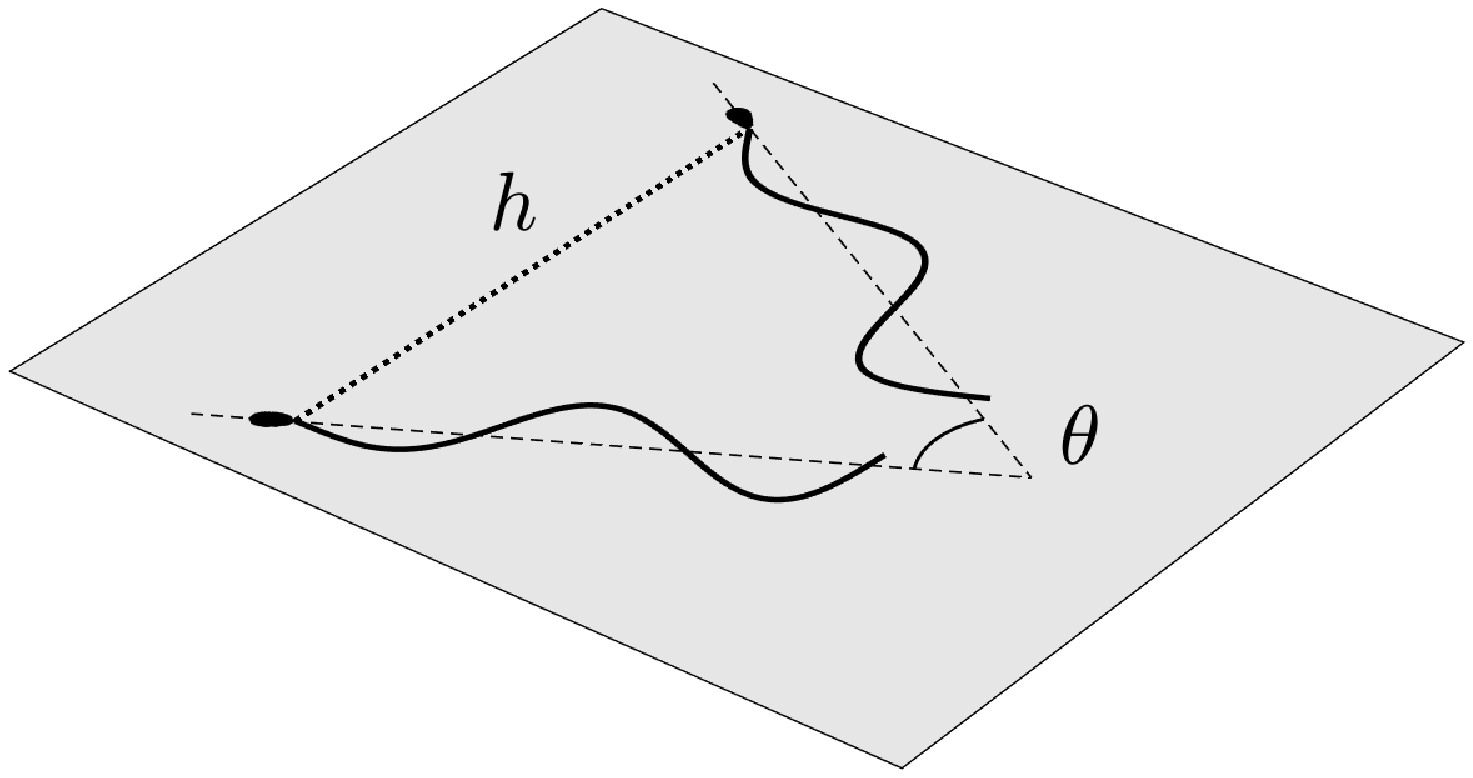}}
  \subfloat[\label{fig:methods:configurations:above-below}]{%
    \includegraphics[width = 0.4\textwidth]{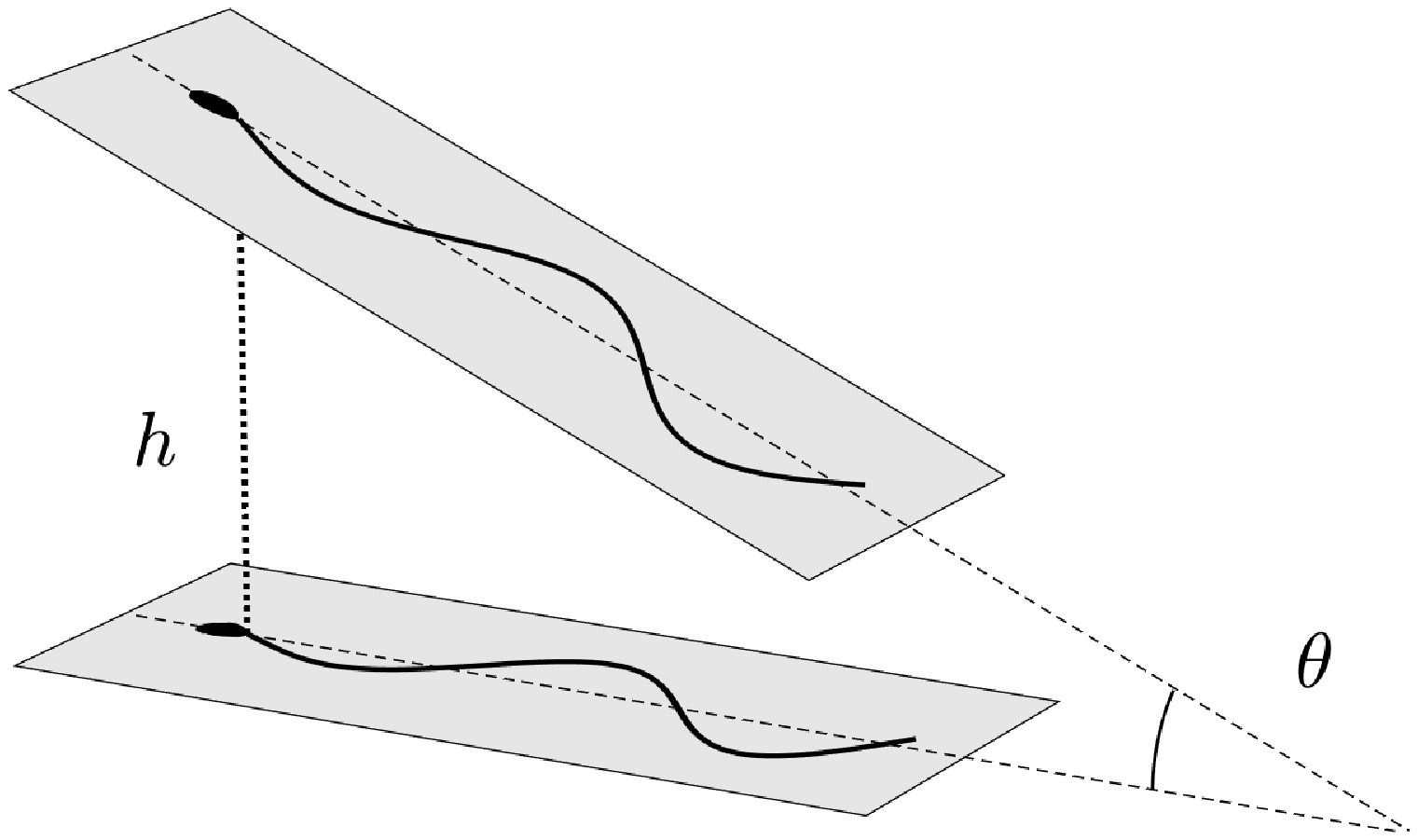}}
  \caption{\label{fig:methods:configurations} Schematic of virtual
  spermatozoa configurations and accompanying parameterizations, with
  free-space headings being shown as dashed lines and their separation
  $\sep{}$ represented by a dotted line.
  \protect\subref{fig:methods:configurations:side-by-side} Swimmers share a
  common beat plane in the side-by-side configuration, with $\heading{}$
  defined as the signed planar angle between their free-space headings.
  \protect\subref{fig:methods:configurations:above-below} In the above-below
  configuration the swimmers' relative heading is defined as the signed
  angle between their beat planes, as shown, where we are making the
  approximation that the relative motion of the swimmers is expressible
  solely by the quantities $\sep{}$ and $\heading{}$, justified a posteriori
  in \cref{sec:results:timescales,sec:results:stability:above-below}. In
  both configurations $\heading{}=0$ corresponds to the virtual spermatozoa
  sharing both beat plane orientation and free-space heading, and we adopt
  the sign convention such that $\heading{}>0$ denotes spermatozoa facing
  away from one another.}
\end{figure}

The interactions of swimming spermatozoa with synchronized flagellar beats
naively form a 13-dimensional system, with 6 scalar quantities representing
the location and orientation of each individual swimmer, in addition to the
phase of the shared flagellar beat. Consideration of only their relative
motion eliminates 6 degrees of freedom, though full exploration of even this
reduced system is computationally prohibitive. We will consider the motion of
swimmers \changed{initiated in one of two} configurations, termed
\emph{side-by-side} and \emph{above-below}, similar to the work of
\citet{Llopis2013} and each of relevance to the wealth of captured
videomicroscopy of spermatozoa, in which nearby swimmers commonly share a beat
plane or lie closely above one another in the plane of imaging, as seen in the
Supplementary Information of \cite{Tung2017,Gadelha2010}. \changed{From these
initial configurations swimmer motion is simulated unconstrained, allowing for
full three-dimensional interactions between swimmers.} In
\cref{fig:methods:configurations:side-by-side} we show a side-by-side
configuration of swimmers, in which the spermatozoa share a beat plane and
their relative motion may be in part described by the separation of their
head-tail junctions, $\sep{}=\abs{\headtail[1]-\headtail[2]}$, and the
relative angle between their free-space headings, $\heading{}$, where we are
neglecting an additional degree of freedom describing their relative in-plane
displacement, to be justified a posteriori given initial conditions with
swimmers directly beside one another.

In \cref{fig:methods:configurations:above-below} we exemplify the above-below
configuration of swimmers, in which virtual spermatozoa beat planes are
parallel to each other except for rotations about the swimmer-fixed
$x_{i2}$-axes. Notably $\heading{}=0$ corresponds to agreement between the
free-space headings of the swimmers, with any rotations within the plane of
beating being neglected. Indeed, we hypothesize that consideration of such
in-plane rotations is of limited physical relevance due to the minimal
magnitude of swimmer interactions in this configuration, a hypothesis that we
will examine in detail in \cref{sec:results:timescales}.

\subsection{Reduction of behavior to an approximate low-dimensional dynamical system}
\label{sec:methods:phase_plane}
In order to systematically capture the complex behaviors of interacting
spermatozoa we simulate the motion of the two swimmers for a single period of
their synchronized flagellar beats, recording their change in separation
$\sep{}$ and relative heading $\heading{}$, initially enforcing that the
swimmers are directly beside or above each other. We note that the dependence
of the relative motion of the swimmers on the initial phase $\psi_0$ is
expected to be minimal, as is confirmed by numerical simulation, and thus we
take $\psi_0\equiv0$ throughout. This effectively reduces the swimming motion
to an approximate two-dimensional dynamical system in $\sep{}$ and
$\heading{}$, \changed{enabling the application of a dynamical systems
framework} as in \citet{Walker2019,Walker2018}. Validity of this methodology
is established by direct comparison with long-time simulations of swimmer
interactions, which highlights in particular that the relative displacement of
swimmers along their respective headings does not qualitatively impact on the
\htheta{} dynamics on relevant timescales from initial conditions of zero
displacement. 

\subsection{Coarse-grained singularity representations of swimmers}
\label{sec:methods:singularity}
We will aim to evaluate the accuracy of a recent \changed{regularized} singularity
representation \cite{Ishimoto2017b,Ishimoto2018b,Chakrabarti2019}, as used by
\citet{Ishimoto2018} to simulate population-level behaviors of spermatozoa
confined to a plane, in particular considering its natural extension to motion
in three spatial dimensions. Here we briefly recapitulate the key principles
of the methodology, and direct the reader to the original publications \cite{Ishimoto2017b,Ishimoto2018b,Chakrabarti2019,Ishimoto2018} for further details.

Given the flow field generated by a swimmer in the absence of any other fluid
boundaries or background flows, principal component analysis is used to
extract high-variance modes of velocity field. The two most-significant modes,
which in this case capture 91.6\% of the variance of the flow field, are
approximated via a small distribution of \changed{regularized} Stokeslets
\cite{Cortez2001}, yielding a low-dimensional representation of the
time-varying flow field associated with a single isolated swimmer that has
been validated to reasonable accuracy even in the presence of a planar
boundary. This representation\changed{, which will be shown graphically in
\cref{sec:results:coarse_grained},} allows efficient simulation of multiple
swimmers in the same domain, with the velocity of any individual swimmer being
determined via force and torque-balance equations, for instance with the drag
on a given swimmer approximated by the sum of the strengths of its constituent
\changed{regularized} Stokeslets.
\section{Results}
\label{sec:results}
\subsection{Swimming in close proximity is transient for side-by-side
spermatozoa}
\label{sec:results:stability:side-by-side}
\begin{figure}[ht]
  \centering
  \subfloat[\label{fig:results:stability:side-by-side:trace}]{%
    \includegraphics[width = 0.44\textwidth]{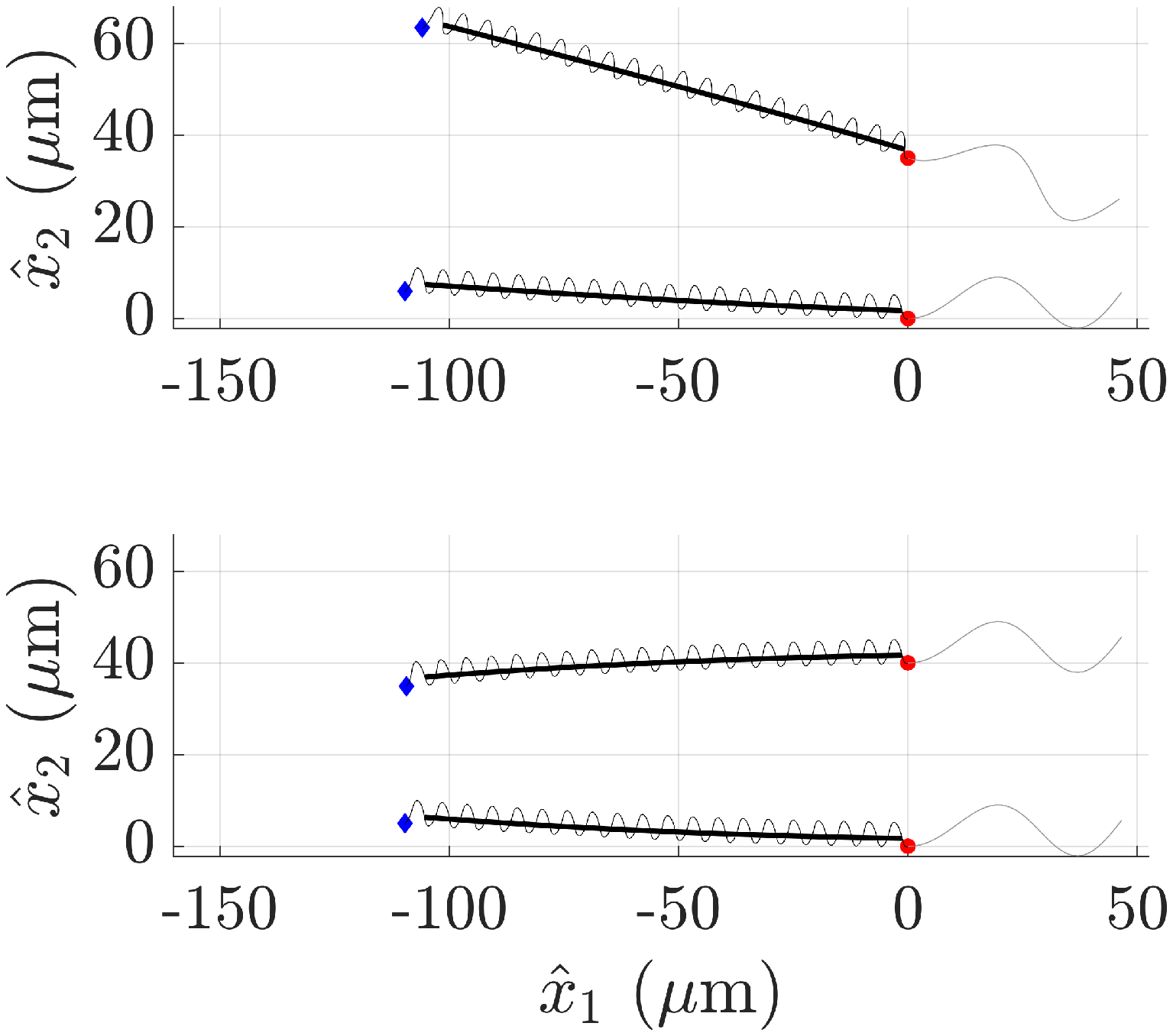} }
  \subfloat[\label{fig:results:stability:side-by-side:phase-plane}]{
    \includegraphics[width = 0.4\textwidth]{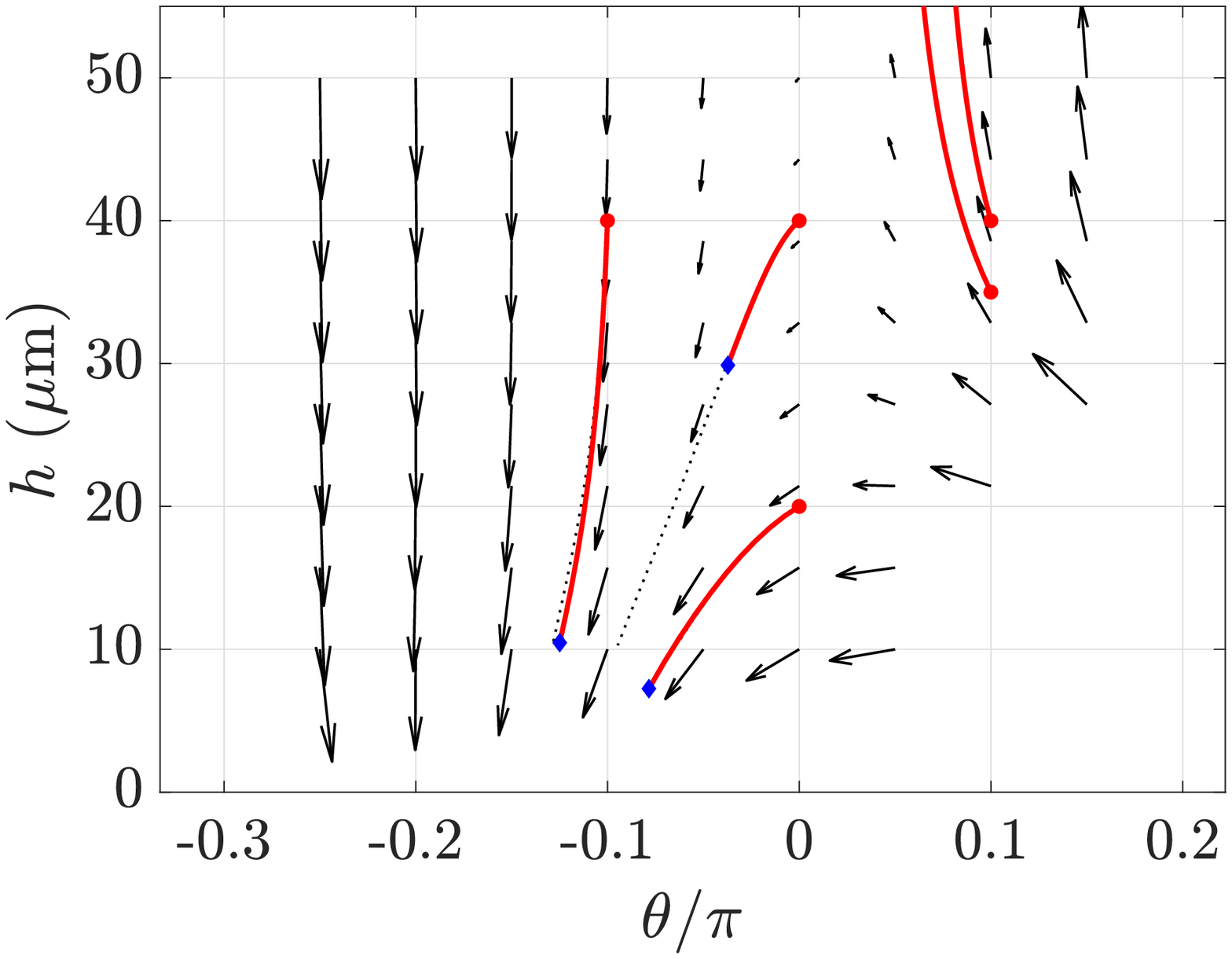}}
  \caption{\label{fig:results:stability:side-by-side}The transient dynamics
  of side-by-side swimming, with trajectory start and endpoints are shown as
  red circles and blue diamonds respectively.
  \protect\subref{fig:results:stability:side-by-side:trace} Each shown as thin
  curves are the positions of the head-tail junctions $\headtail[i]$ of a pair
  of spermatozoa in the laboratory frame $\hat{x}_1\hat{x}_2\hat{x}_3$ for
  side-by-side swimmers with initial conditions $(\sep{},\heading{}) =
  (35\si{\um},0.1\pi)$ and $(\sep{},\heading{}) = (40\si{\um},0)$, for the
  upper and lower panels respectively, and simulated over 20 periods of the
  flagellar beat. Averaged swimmer paths are shown as heavy black curves,
  highlighting separation in the upper panel and convergence of the swimmers
  in the lower panel. \changed{Shown in gray are the swimmers in their
  original configurations, with the cell bodies omitted here for clarity.}
  \protect\subref{fig:results:stability:side-by-side:phase-plane} Pairwise
  dynamics in the \htheta{} space, with the projections of sample long-time
  simulations, typically simulated for 20 periods of the flagellar beat,
  shown as solid red curves. Solutions of the approximate plane autonomous
  system are shown dotted with the same initial conditions as the long-time
  simulations, demonstrating very good agreement between the full and
  reduced systems, serving as a posteriori validation of the reduction to
  \htheta{} space. Further, the dynamics as predicted by the phase plane
  exhibit no stable pairwise swimming, merely the behaviors of eventual
  collision and separation off into the bulk, though sperm initially heading
  away from one another are predicted to sometimes still be on a path that
  will result in eventual collision. Color online.}
\end{figure}

The long-time motion of two virtual spermatozoa in close proximity in a
side-by-side configuration was simulated using the boundary element method
over numerous beat cycles, where both swimmers are equipped with the same
in-phase flagellar beat and are swimming within the same plane. Simulations
from various initial configurations demonstrated that nearby swimming in a
side-by-side configuration is transient, with the swimmers either colliding
with or deflecting away from one another. Initial separations of the head-tail
junctions were sampled in the range $h\in[2.5\si{\um},50\si{\um}]$, with the
relative angle of free-space progression $\heading{}$ ranging from $-\pi/4$ to
$\pi/4$. Typical traces showing the location of the head-tail junction of each
swimmer over time are presented in
\cref{fig:results:stability:side-by-side:trace} for side-by-side swimming,
demonstrating collision and deflection of the swimmers that is dependent on
their relative initial configurations.

In order to systematically classify these behaviors of collision and
deflection, we proceed by considering the approximate dynamical system as
described in \cref{sec:methods:phase_plane}. The resulting phase plane is
shown in \cref{fig:results:stability:side-by-side:phase-plane}, along with
projections of long-time simulations of unconstrained swimmer motion,
demonstrating very good agreement between swimmer motion and the approximate
reduction to a plane autonomous system. In particular, this agreement
validates a posteriori the use of such a phase plane in studying the dynamics
of a swimming pair in this configuration, and highlights that any relative
motion of the swimmers along their respective axes of progression does not
significantly alter their behaviors on reasonable timescales from these
initial conditions; indeed, the sample long-time trajectories shown in
\cref{fig:results:stability:side-by-side:trace} exemplify that such
displacement is minimal. Absent from the plane autonomous system is a stable
swimming configuration or limit cycle, meaning that trajectories either result
in collision or represent swimmers separating off into the bulk, and thus
pairwise side-by-side swimmer motion is transient. Remarkably, we observe from
the phase plane that even swimmers initially facing away from one another may
eventually collide as the result of proximity-induced reorientation, with the
magnitude of this effect decreasing as the separation \changed{between their
head-tail junctions} $\sep{}$ increases.

\subsection{Above-below swimming dynamics approach a stable spiral}
\label{sec:results:stability:above-below}
\begin{figure}[ht]
  \centering
  \includegraphics[width = 0.5\textwidth]{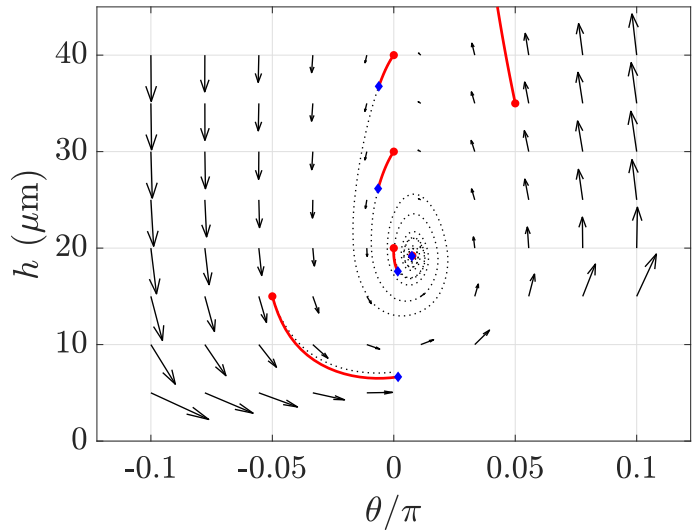}
  \caption{\label{fig:results:stability:above-below}The approximate dynamics of
  above-below swimming, highlighting a pairwise swimming mode that
  corresponds to stable long-time motion in close proximity, represented
  here by the existence of an attracting stable spiral around
  $\sep{}\approx19.1\si{\um}$, $\heading{}\approx0.0076\pi$. Projections of
  long-time simulations, computed for 20 periods of the flagellar beat and
  serving as a posteriori validation of the phase plane, are shown as red
  curves, with start and endpoints displayed as red circles and blue
  diamonds respectively. In particular, a long-time simulation initiated
  from approximately the stationary point of the autonomous system shows
  that swimmers indeed undergo negligible relative motion, verifying the
  existence of a stable swimming configuration. Solutions of the approximate
  plane autonomous system are shown dotted with the same initial conditions
  as the long-time simulations, demonstrating very good agreement between
  the full and reduced systems. Trajectories can be seen to spiral inwards,
  approaching the stationary point on timescales on the order of hundreds of
  beat cycles. Color online.}
\end{figure}

Having examined the interactions of swimmers that share a common beat plane,
we repeat the phase plane quantification of pairwise swimmer dynamics for
virtual spermatozoa in an above-below configuration, with swimmers initially
situated directly above one another. The reduced system, shown in
\cref{fig:results:stability:above-below}, exhibits convergence to a stable
spiral in the \htheta{} space from a wide range of initial configurations in
the phase space, assuming that initially the free space headings of the
swimmers lie within parallel planes and that the virtual spermatozoa begin
located directly above one another. With the central fixed point located at
$(\sep{},\heading{})\approx(19.1\si{\um},0.0076\pi)$, this spiral corresponds
to swimmers slowly oscillating about a configuration where they are oriented
almost parallel to one another, and represents a stable mode of pairwise
swimming in close proximity. The existence of a stationary point in the
\htheta{} space may be explained intuitively by a simple balance
of hydrodynamic attraction between swimmers and their progression along their
free-space headings, noting that swimmers are oriented slightly away from each
other in the stable configuration, though it is a priori unclear that such
mutual attraction should be present at all.

In order to validate the existence of this stable swimming behavior in the
full dynamics, a number of long-time simulations of unconstrained swimmer
motion were performed, with sample projections onto the \htheta{} space shown
as red curves in \cref{fig:results:stability:above-below}. As was the case for
side-by-side swimmers, excellent agreement is observed in general between
trajectories predicted from the plane autonomous system and those projected
from long-time simulations. In particular, a simulation of above-below
swimmers beginning in the steady-state configuration predicted by the reduced
dynamics demonstrates pairwise swimming behavior corresponding to the stable
spiral in the \htheta{} space\changed{, and is robust to perturbations out of
an initial above-below configuration}. Thus, from consideration of a
validated reduction of the dynamics to a plane autonomous system we have
identified a stable form of pairwise swimming in virtual spermatozoa, present
for swimmers in an above-below configuration and corresponding to
approximately-parallel motion. Further, and most remarkably, we have observed
that this mode of behavior is indeed exhibited in long-time simulations of the
full pairwise dynamics.

\subsection{Timescales of pairwise interaction are configuration dependent}
\label{sec:results:timescales}
\begin{figure}[ht]
  \centering
  \includegraphics[width=0.4\textwidth]{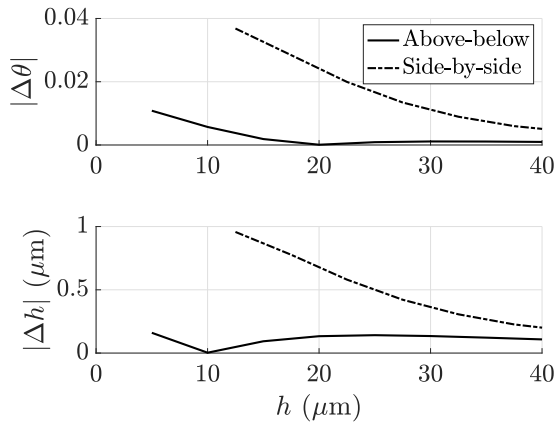}
  \caption{\label{fig:results:timescale}Change in swimmer separation and
  relative heading over a single beating period as a function of initial
  separation $\sep{}$, denoted $\Delta\sep{}$ and $\Delta\heading{}$
  respectively. Initially the swimmers are situated directly above or beside
  one another, and with $\heading{}=0$. Above-below swimmers, shown as solid
  curves, can be seen to undergo reduced changes in relative heading and
  separation when compared to side-by-side swimmers, shown dot-dashed,
  highlighting a disparity in the timescales of relative motion between
  swimmer configurations.}
\end{figure}

Further to enabling an analysis of the overall behaviors of swimming
spermatozoa, identifying collision, deflection or stable relative motion, the
long-time simulations and approximate phase plane computations of
\cref{sec:results:stability:side-by-side,sec:results:stability:above-below}
reveal a disparity in timescales between swimmer configurations. Whilst this
may be deduced from detailed consideration of the phase planes of
\cref{fig:results:stability:side-by-side,fig:results:stability:above-below},
this is most-clearly exemplified by focusing on the relative motion of
swimmers in each configuration with $\theta=0$ initially. In
\cref{fig:results:timescale} we report the change in separation
\changed{between their head-tail junctions along with the difference in}
relative heading over a single beat period as a function of initial
separation, as used to generate the profiles of the reduced systems in
\cref{sec:results:stability:side-by-side,sec:results:stability:above-below}.
From this we note that the change in both swimmer separation and relative
heading over a beat period are significantly greater for side-by-side swimming
than for swimmers in an above-below configuration, with the change in heading
in particular being approximately an order of magnitude less for above-below
swimmers. Hence we conclude that the hydrodynamic effects of proximity on
swimming behavior are most prominent for virtual spermatozoa that share a
plane of beating, in contrast to the relatively weak effects experienced by
swimmers in an above-below configuration, highlighting a stark anisotropy in
the interactions of neighboring swimmers.

In particular, the reduced magnitude of swimmer interactions in the
above-below configuration justifies a posteriori our consideration of
above-below swimming only where there is no relative rotation within the
swimmers' planes of beating. Indeed, were there to be such a relative
rotation, virtual spermatozoa may be expected to simply pass over each other
and separate with little observable interaction, owing to the noted relative
subdominance of out-of-plane hydrodynamic interactions in this configuration.
 
\subsection{Effects of proximity on swimming speed are anisotropic and commonly
inhibit progression}
\label{sec:results:speed}
With swimming stable in above-below configurations and transient over long
timescales for side-by-side swimmers, we consider in more detail the effects
of low cell separation on swimmer progression. In particular, for swimmer
pairs in each of the side-by-side and above-below configurations we compute
the average linear velocity of the swimmers over a single beating period,
where as above the flagellar beats are assumed to be in synchrony.

\begin{figure}[ht]
  \centering
  \includegraphics[width = 0.4\textwidth]{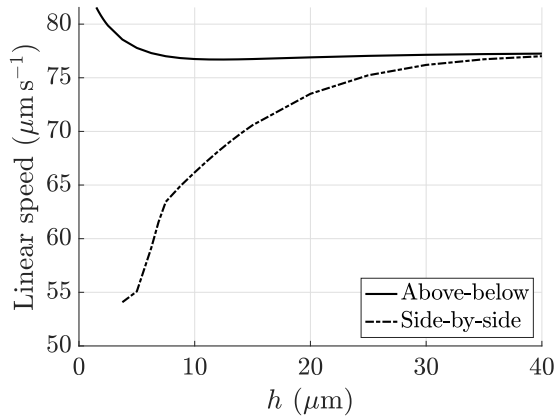}
  \caption{\label{fig:results:linear_velocity}The linear velocity of a
  swimmer over a single beating period as a function of separation, $h$, in
  both above-below and side-by-side configurations. With swimmers initially
  directly above/beside one another, and shown here with $\heading{}=0$,
  observed for both configurations is a reduction in swimming speed with
  separation \changed{between swimmer head-tail junctions}, though the
  magnitude of such reduction is notably greater in the case of side-by-side
  swimming than for above-below swimmers, the change in the latter being
  difficult to observe on the scale of the above plot. At further-reduced
  separations, below approximately 10\si{\um}, remarkably, virtual spermatozoa
  in an above-below configuration can be seen to increase in speed, surpassing
  even their free-space swimming speed of approximately 77\si{\um\per\s}.}
\end{figure}

Considering first the case of side-by-side swimming, the swimming speed being
shown as a dot-dashed curve in \cref{fig:results:linear_velocity}, we see that
a reduction in swimmer separation yields a substantially-reduced average
swimming speed, with each swimmer inhibiting the progression of the other. The
same general trend may be observed for virtual sperm in an above-below
configuration, with speed shown as a solid curve, though reductions in
swimming speed are much less in this instance. However, in contrast to
side-by-side swimming, in the above-below configuration we observe a notable
increase in linear speed when they are in very close proximity of one another,
with separations of less than approximately 6.25\si{\um} giving rise to
average speeds greater even than those attained by lone swimmers in the bulk.
The magnitude of this effect increases further as separation is reduced, with
an increase of around 4\si{\um\per\s} over the bulk speed when at a separation
of 1.5\si{\um}\changed{, with swimmers nearing contact in this configuration}.
Thus, in addition to the aforementioned anisotropy of timescales, the effects
of proximity on the speed of progression of the virtual spermatozoa are also
configuration-dependent, in qualitative agreement with the results of
\citet{Llopis2013} though here observed accounting for the swimmer body and an
idealized spermatozoan beating pattern. We observe that side-by-side swimming
invariably results in reduced swimming speed, whilst close virtual spermatozoa
in an above-below configuration have increased linear speed compared to both
side-by-side swimmers and even lone cells in the bulk.

\subsection{Efficacy of PCA-derived singularity representations is limited in three-dimensional interactions}
\label{sec:results:coarse_grained}

\begin{figure}[b]
  \centering
  \subfloat[\label{fig:results:flow_field:real}]{%
    \includegraphics[width = 0.3\textwidth]{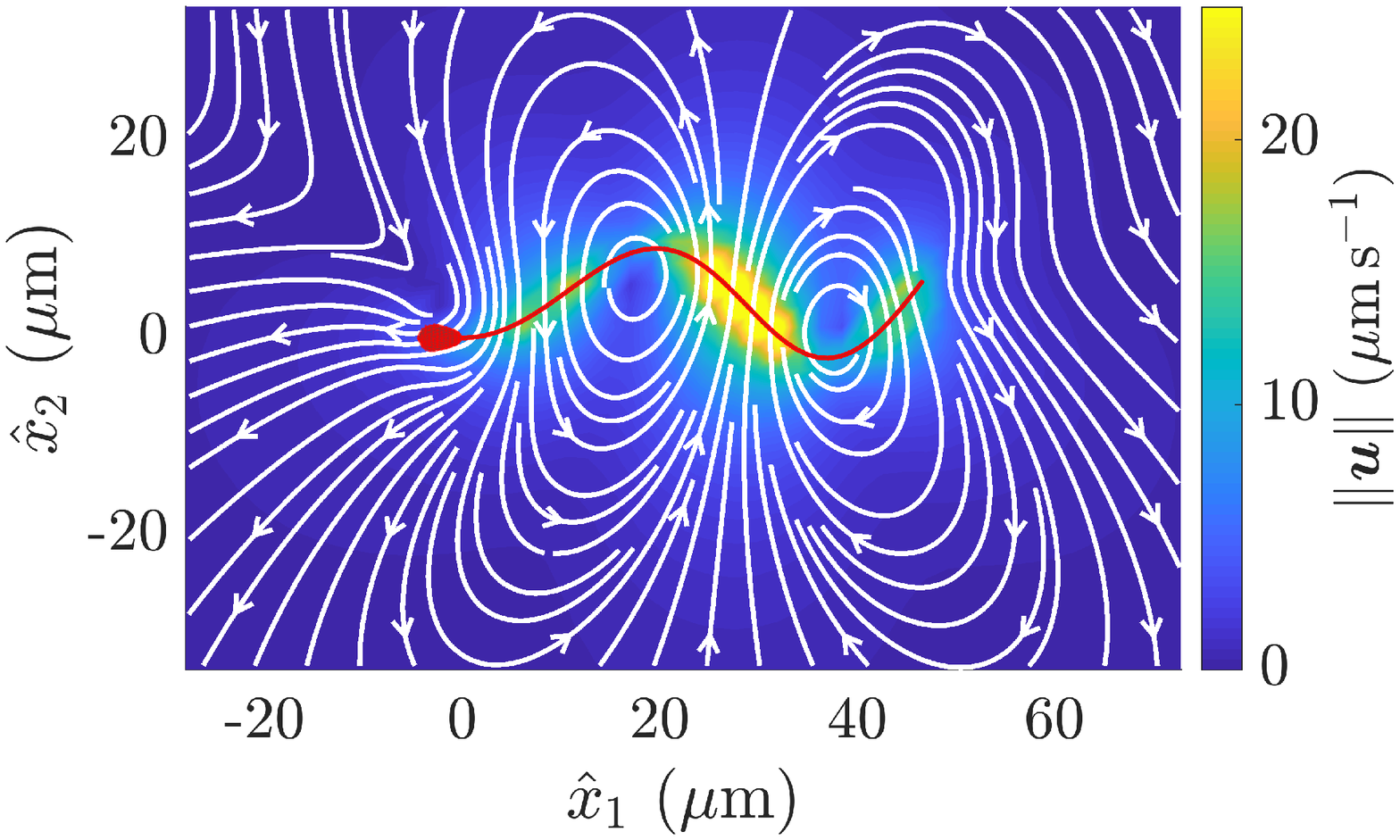}}
  \subfloat[\label{fig:results:flow_field:singularity}]{%
    \includegraphics[width = 0.3\textwidth]{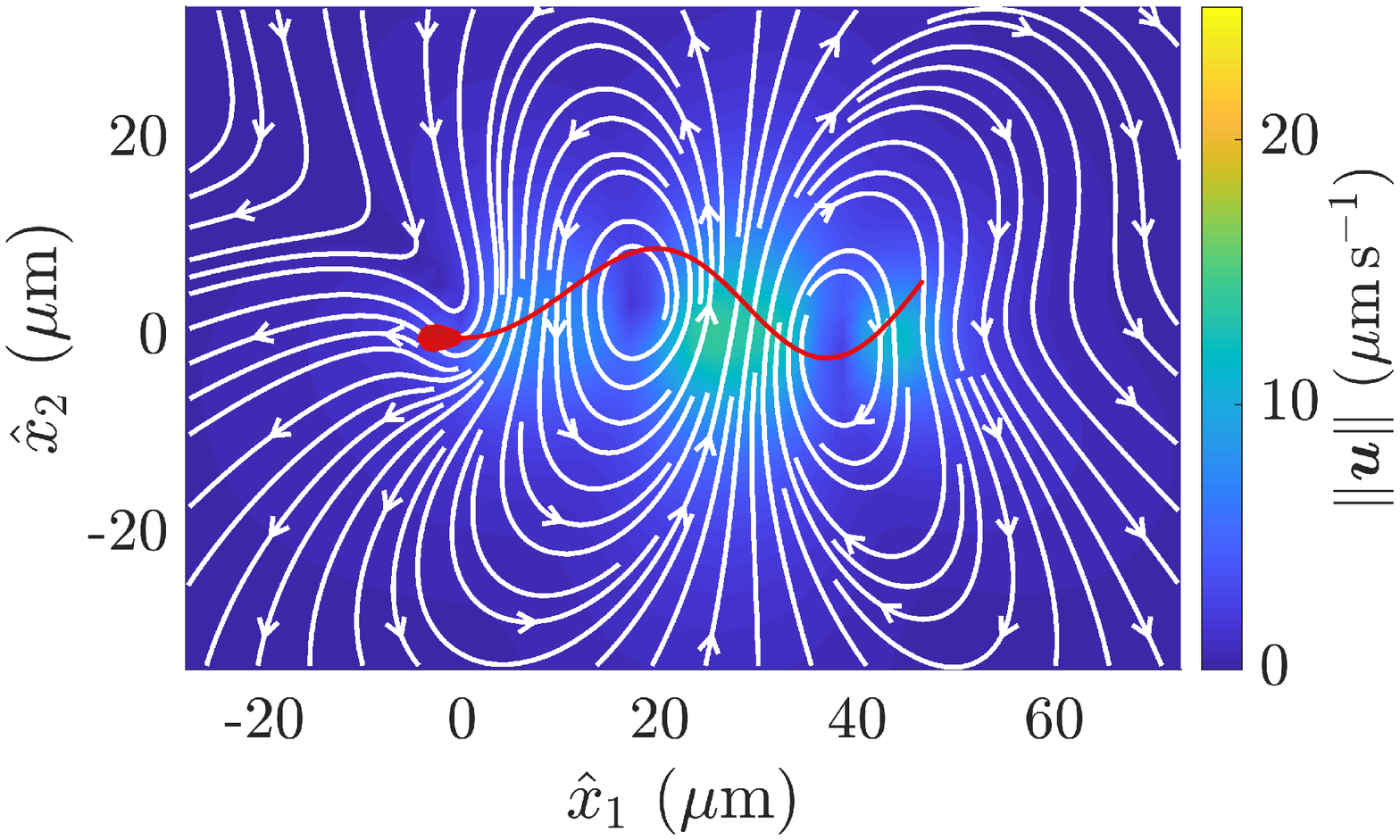}}
  \subfloat[\label{fig:results:flow_field:modes}]{%
    \includegraphics[width = 0.3\textwidth]{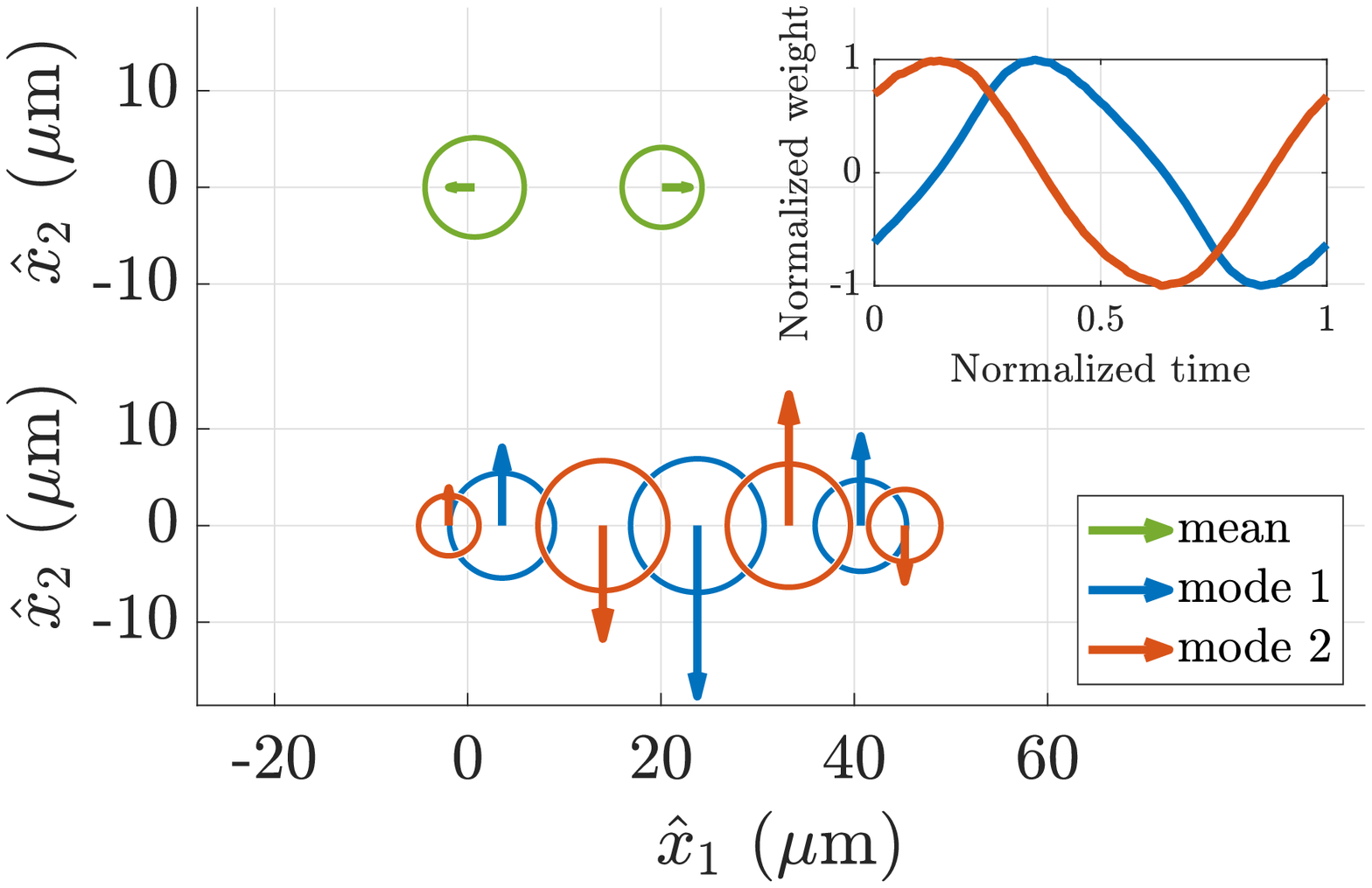}}
  \caption{\label{fig:results:flow_field}Instantaneous flow field in the
  plane of beating for an individual virtual spermatozoon in the laboratory
  frame $\hat{x}_1\hat{x}_2\hat{x}_3$, with the swimmer shown in red and
  situated at the origin of the laboratory frame. The flow field as computed
  by the boundary element method is given in
  \protect\subref{fig:results:flow_field:real}, with the corresponding
  \changed{regularized} singularity representation flow field shown in
  \protect\subref{fig:results:flow_field:singularity}.
  \changed{\protect\subref{fig:results:flow_field:modes} An illustration of
  the low-dimensional \changed{regularized} singularity representation,
  comprising of the lowest two PCA modes and the mean over a beating period,
  with arrow size and direction corresponding to the strength and direction of
  the regularized Stokeslets. The circle size corresponds to the
  regularization parameter taken for each regularized Stokeslet, and inset is
  the normalized weighting of each of the PCA modes over a single beating
  period. Each regularized Stokeslet lies on the line $\hat{x}_2=0$ by the
  symmetry of the flagellar beat, with those corresponding to the mean flow
  here being shown offset for visual clarity.} Qualitative details of the
  instantaneous flow field can be seen to be well-approximated by the
  low-order regularized singularity representation. The magnitude of the flow
  fields are given by the intensity of the shading, with projected streamlines
  shown in white. Color online.}
\end{figure}

We simulate the motion of a single swimmer in an unbounded domain to determine
the swimmer's characteristic time-varying flow field, and form the PCA-derived
singularity representation of the swimmer as described in
\cref{sec:methods:singularity}, with \changed{both flow fields and the
accompanying \changed{regularized} singularity representation} being shown in
\cref{fig:results:flow_field}. Simulating pairwise swimmer motion as in
\citet{Ishimoto2018}, we proceed to evaluate the accuracy of using such
singularity representations for the simulation of swimmer interaction by
repeating the analysis of
\cref{sec:results:stability:side-by-side,sec:results:stability:above-below,sec:results:speed}
using this approximate framework.

\begin{figure}[ht]
  \centering
  \subfloat[\label{fig:results:PCA:linear_velocity}]{%
    \includegraphics[width = 0.3\textwidth]{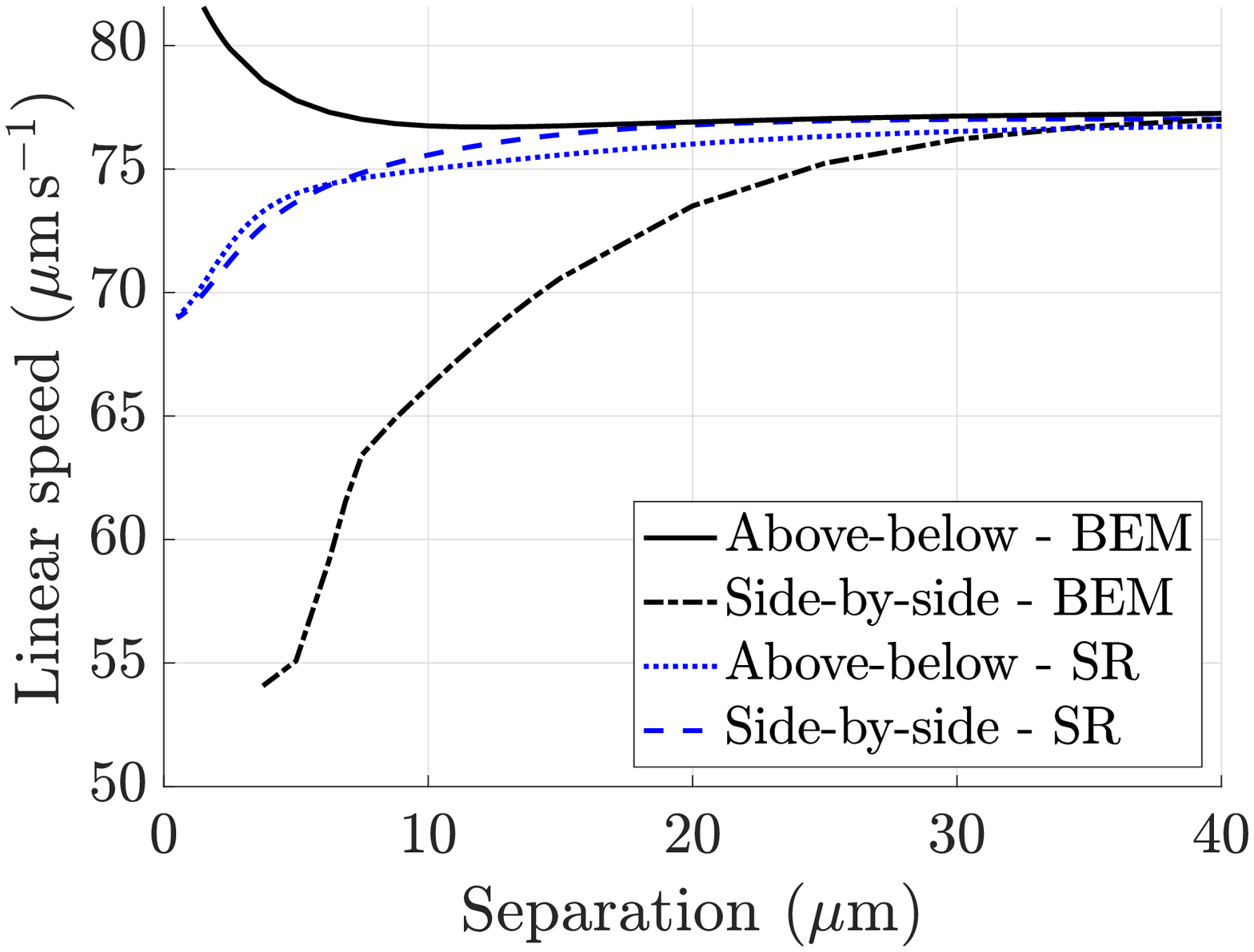}}
  \subfloat[\label{fig:results:PCA:side-by-side-phase}]{
    \includegraphics[width = 0.3\textwidth]{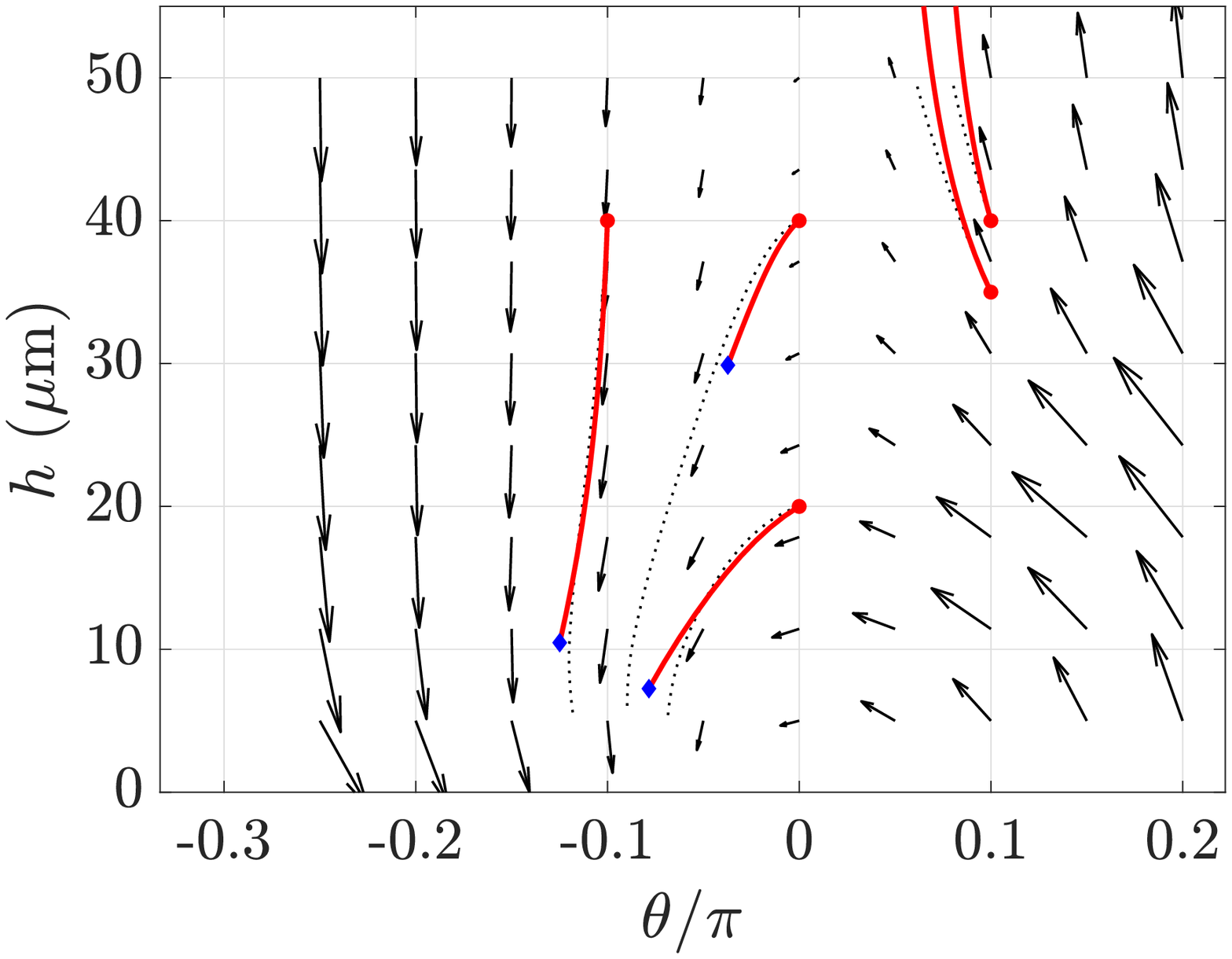}}
  \subfloat[\label{fig:results:PCA:above-below-phase}]{
    \includegraphics[width = 0.3\textwidth]{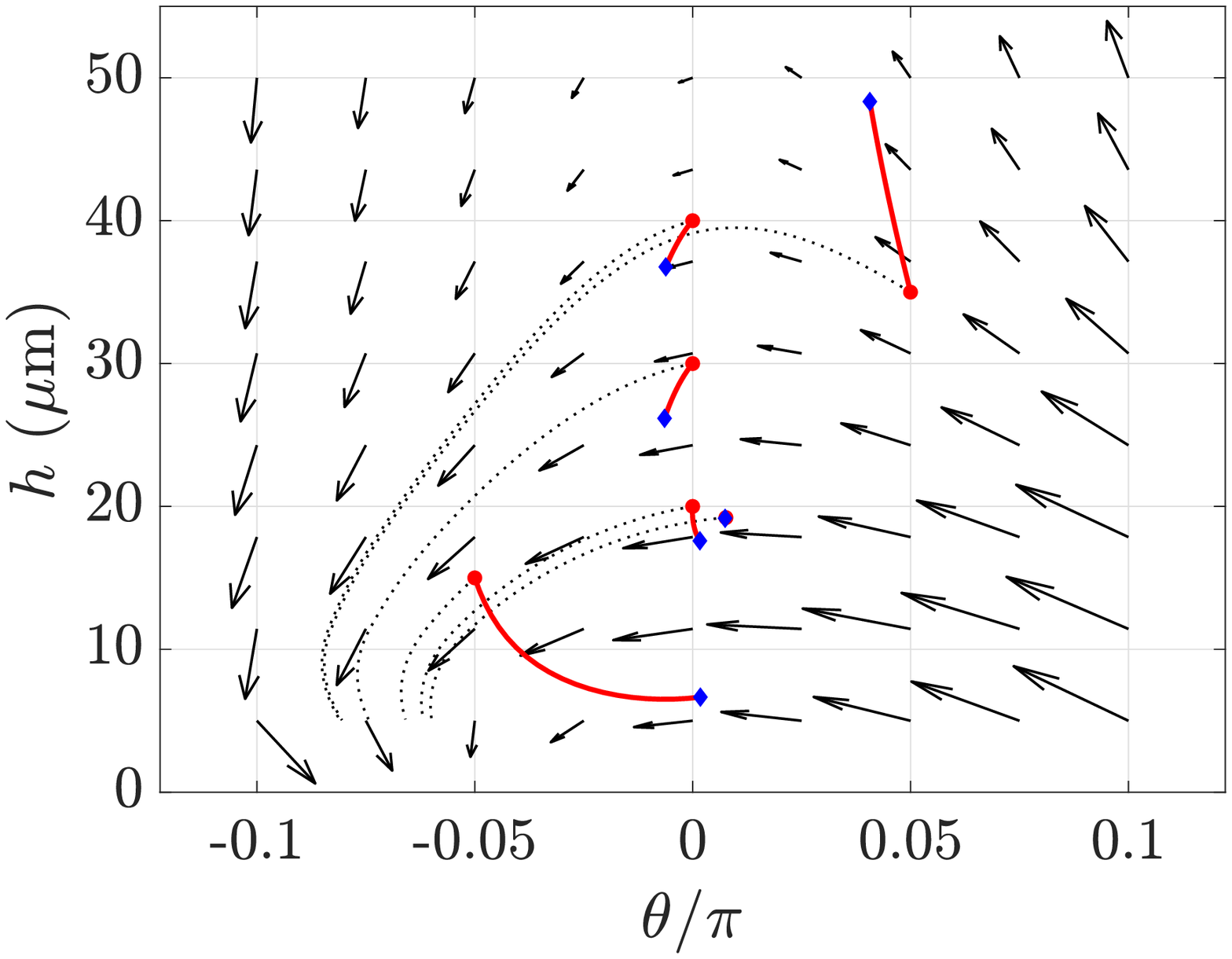}}
  \caption{\label{fig:results:PCA}Comparison of swimmer behaviors as
  predicted by the boundary element method and PCA-derived singularity
  representations. \protect\subref{fig:results:PCA:linear_velocity} The
  PCA-derived singularity representations (SR), shown here as blue curves,
  agree qualitatively with the boundary element method (BEM) predictions for
  linear swimming speed in the case of side-by-side swimming, though no such
  agreement is present in close proximity for swimmers in an above-below
  configuration.
  \protect\subref{fig:results:PCA:side-by-side-phase}--\protect\subref{fig:results:PCA:above-below-phase} Evolution in \htheta{}
  space for \protect\subref{fig:results:PCA:side-by-side-phase} side-by-side
  and \protect\subref{fig:results:PCA:above-below-phase} above-below
  swimmers as predicted by singularity representations, shown as black
  arrows, with trajectories of the plane autonomous system shown as dotted
  black curves. Superimposed as solid red curves are projections of boundary
  element simulations onto the reduced space, with start and endpoints shown
  as red circles and blue diamonds respectively. We note good agreement in
  the case of side-by-side swimming, though there is little correspondence
  between the predictions of the boundary element method and the singularity
  representations for swimmers in an above-below configuration. Levels of
  agreement are further evidenced by comparison with
  \cref{fig:results:stability:side-by-side,fig:results:stability:above-below}.
  Color online.}
\end{figure}

Firstly, we consider the predictions of this methodology with regards to
separation-dependent swimming speed, reported above using the boundary element
method as strongly dependent on swimmer configuration. With the results shown
in \cref{fig:results:PCA:linear_velocity} as blue curves, we observe that the
singularity representation is unable to capture even the qualitative effects
of swimmer proximity on swimming speed in the above-below configuration,
highlighting a subtlety in the mechanism by which swimmers may mutually
benefit from pairwise swimming. However, for side-by-side swimmers the
concordance is fair, particularly as swimmer separation increases, suggesting
a potential efficacy of this simple singularity representation in describing
all but the closest of swimmer interactions in this configuration.
Unsurprisingly, at large separations the cases of above-below and side-by-side
converge to the free-space swimming speed of our virtual spermatozoa, as do
the results of both the boundary element method and singularity representation
computations, with swimmer-swimmer interactions decaying in magnitude as
separation increases.

With reasonable agreement in swimming speed seen for side-by-side swimmers, we
examine the long-time behaviors of pairwise swimmers in this configuration as
in \cref{sec:results:stability:side-by-side}, instead using the PCA-derived
singularity representations, simulating swimmer motion over a single beating
period from various initial orientations and separations and forming the
reduced dynamical system of \cref{sec:methods:phase_plane}. Shown in
\cref{fig:results:PCA:side-by-side-phase} are the dynamics in the
\htheta{} space, with projections of long-time boundary element simulations
superimposed as red curves for direct comparison. We observe very good
qualitative agreement in overall swimmer behavior between the predictions of
the boundary element method and the singularity representations, consistent
with that noted for linear swimming speeds above. This consensus suggests that
the singularity-based methodology provides a description of side-by-side
dynamics that is capable of capturing qualitative swimmer behavior, and thus
may be widely applicable to studies of large scale population interactions.

Now considering swimmers in an above-below configuration and repeating the
construction of the approximate plane autonomous system as above, by drawing
comparison with \cref{fig:results:stability:above-below} we see that the
singularity representation fails to capture the characteristics of the motion
as predicted by the boundary element method. Most notably, no stable swimming
is present as a mode of pairwise swimming, with swimmers moving off into the
bulk or colliding. Further, the overall dynamics of the reduced system
computed via the singularity representation are not dissimilar to those
predicted for side-by-side swimmers, suggesting that the anisotropy seen in
boundary element simulations of our virtual spermatozoa is not captured by the
low-order singularity representation.

Overall, we have seen good agreement between methodologies for swimmers in a
side-by-side configuration, but little coherence in predicted behaviors for
virtual spermatozoa in above-below configurations. Hence we may conclude that
the PCA-derived singularity representations should not be relied upon in their
current form for the simulation of three-dimensional interactions in
generality, though provide remarkable qualitative agreement with
computationally-expensive high-accuracy methods for planar swimmers sharing a
beating plane.

\section{Discussion}
\label{sec:discussion}
This study has considered the motion of two virtual spermatozoa in close
proximity, restricted to planar beating and to two configurations of interest:
side-by-side and above-below swimming. In doing so we have identified a
significant anisotropy in their relative motion, not simply limited to
quantitative effects. On examining the linear swimming speed attained by our
virtual, synchronized swimmers over a single beat period, we have observed
that side-by-side spermatozoa make less progress along their respective
headings than they would achieve in isolation. Whilst the same albeit
substantially weaker general trend may be noted in above-below swimmers at
separations greater than approximately 10\si{\um}, for reduced separations we
note a remarkable increase in swimming speed, above that attained by lone
swimmers in the bulk. We hypothesize that the pairwise attachment of
spermatozoa in an above-below configuration may afford them a competitive
advantage over non-bound individuals, with proximity increasing their speed of
progression in-line with the predictions of our computational model. In
agreement with the predictions of the bead-and-spring model of
\citet{Llopis2013}, our findings however are in part distinct from the similar
computational results of \citet{Olson2015} concerning side-by-side elastic
filaments, which predict that swimming speeds increase with proximity for
planar such filaments. This disparity in conclusions suggests that the
inclusion of swimmer head geometry may have significant impact on the
predicted qualitative behaviors of the swimming pair, with care being
necessitated in future theoretical works to accurately capture such effects.

Further anisotropy is present in the timescales of relative motion. We have
seen that swimmers in an above-below scenario undergo greatly-reduced changes
to their configurations over a beating cycle when compared to the motion of
swimmers sharing a beat plane. This slow evolution in relative position and
direction for swimmers situated in an above-below configuration appears to be
in approximate concordance with the general experimental observation that
motile spermatozoa frequently appear to cross one another with little visible
interaction, examples being found in the Supplementary Movies accompanying the
recent works of \citet{Tung2017} and \citet{Gadelha2010}, concerning bovine
and human spermatozoa respectively. \changed{Further evidenced in the
Supplementary Movies of \citet{Tung2017}, and additionally in those
accompanying the work of \citet{Ishimoto2017b}, are the limited effects of
noise on swimming behaviors, with Brownian noise typically insignificant on
the lengthscale of spermatozoa and synchronization persisting over reasonable
timescales. However, thorough exploration of the impacts of intra-spermatozoon
variation on the pairwise behaviors reported in this work represents
significant and pertinent future study, in addition to detailed investigations
into the effects of background flows and confined geometries.}

We have seen that the pairwise dynamics of virtual spermatozoa in side-by-side
and above-below configurations may be described with notable accuracy by a
plane autonomous system, motivating potential future consideration of pairwise
interactions of further flagellated microswimmers in this way, extending in
principle the approach of \citet{Ishikawa2006} though here restricted to
particular swimmer configurations due to the absence of symmetry. Most
significantly, analysis of these reduced systems revealed a significant
disparity between the long-time dynamics of above-below and side-by-side
swimming, with the latter only exhibiting behaviors of collision and
separation off into the bulk fluid. Consideration of above-below dynamics led
to the identification of a remarkable mode of stable pairwise swimming in
virtual spermatozoa, corresponding to a stable spiral in the plane autonomous
system and whose existence was confirmed in silico by long-time simulation of
a swimming pair. These behaviors are in partial agreement with general
classical conclusions drawn from representing swimmers as a force dipole,
valid in the far-field for lone swimmers, which predicts that pushers like our
virtual spermatozoa will in general swim facing one another \cite{Lauga2009},
evidenced here by the noted reorientation of side-by-side swimmers. We have
further seen that in fact near-field effects are significant in determining
swimmer interactions, and indeed give rise to a behavioral anisotropy that is
not a feature of this most-simple of far-field swimmer representations, with
stable swimming only being observed here for above-below spermatozoa. The
similar study of \citet{Llopis2013} concerning active bead chains found that
both above-below and side-by-side swimmers ultimately swim away from one
another, further suggesting that the predicted interactions of nearby swimmers
are sensitive to the inclusion or neglect of complex cellular morphology.

The existence of a stable swimming mode in silico raises the hypothesis that
in vitro spermatozoa may swim stably as pairs even without adhering to one
another, though the potentially-complex effects of variation between
individual swimmers are unclear and require detailed consideration. Indeed,
throughout we have considered spermatozoa equipped with a synchronized
kinematic description of their flagellar beating. Relaxation of this
assumption necessitates careful treatment of the governing elastohydrodynamics
and the internal molecular motor dynamics in order to fully capture the
complex interactions between flow field and flagella, and in particular is
likely to be a topic of significant future study, as is the study of bound
swimming sperm in the context of rodent sperm trains
\cite{Moore2002,Fisher2010}.

Having simulated the behaviors of nearby swimmers using a high-accuracy
boundary element method, we explored the qualitative accuracy of a simplified,
coarse-grained approach for multi-swimmer simulation, as implemented by
\citet{Ishimoto2018}. We seen that this simplified approach is sufficient to
capture qualitative features of side-by-side swimming, supporting its use in
such circumstances. However, poor agreement between this methodology and the
boundary element calculations was noted for above-below swimmers, with the
singularity representation failing to reproduce the anisotropy associated with
our planar beaters. This highlights the complexity of the near-field
interactions of these flagellated swimmers, which are more intricate than is
captured by our intuitive physical interpretation of the flow field via such a
low-dimensional singularity representation\changed{, though differences
between methodologies in predicted flow-field magnitude in the very near
field, as seen in \cref{fig:results:flow_field}, may also contribute to the
observed behavioral disparity}. As the singularity method utilizes only
\changed{regularized} Stokeslets in its low-dimensional representations of the swimmers,
we hypothesize that the inclusion of higher-order singularities may render
enhanced anisotropy and improve its efficacy for simulating the motion of
out-of plane swimming pairs, and is a potential direction for the refinement
of this methodology. Without such an augmentation, the methodology of
\citet{Ishimoto2018} thus appears unreliable for three-dimensional pairwise
interactions in general, though is suitably accurate for in-plane interactions
whilst affording significant computational simplicity and scalability to
population-level models.

In summary, we have examined the behavioral consequences of hydrodynamic
interactions between two synchronized flagellated swimmers moving in three
dimensions, identifying a range of complex and subtle pairwise behaviors in
virtual spermatozoa. Long-time simulations using high-accuracy boundary
element simulations revealed diverse behavioral modes that are
configuration-dependent, with side-by-side swimming invariably resulting in
collision or separation into the bulk. Throughout we have seen evidenced an
anisotropy in the dynamics of nearby swimming, with proximity able to effect
both increases and decreases in swimming speed depending on swimmer
configuration, though such subtleties may not be captured by simulations using
only low-order singularity representations of swimmers. Further, we have found
a disparity in the timescales of relative motion between above-below and
side-by-side swimming, and have also noted that reductions to low-dimensional
autonomous systems can well-describe pairwise swimmer dynamics. Finally, from
analysis of such systems we have identified a remarkable swimming mode in
synchronized virtual spermatozoa, one corresponding to stable pairwise
swimming.

\section*{Acknowledgements}
B.J.W.\ is supported by the UK Engineering and Physical Sciences Research
Council (EPSRC), grant EP/N509711/1. K.I. is supported by JSPS Overseas
Research Fellowship (29-0146), MEXT Leading Initiative for Excellent Young
Researchers (LEADER), and JSPS KAKENHI Grant Number JP18K13456.

\appendix*
\section{Parameterization of the virtual spermatozoon head}
\label{app:head_parameterisation}
With respect to the swimmer-fixed reference frame of spermatozoon
$i\in\{1,2\}$, denoted $x_{i1}x_{i2}x_{i3}$, as defined in the main text, we
parameterize the swimmer head by length parameter $\xi\in[0,L_{\text{h}}]$ and
$\eta\in[0,2\pi)$, where $L_h$ denotes the length of the virtual spermatozoon
head, taken to be 4.5\si{\um}. Adopting the idealized geometry of
\citet{Smith2009}, the surface of the spermatozoon head is parameterized by
$(x_{i1},x_{i2},x_{i3}) = (-\xi,\changed{r\sin\eta,r\cos\eta})$,  where $r$ is given in
terms of $\xi$ and $\eta$ by
\begin{align}
  r^2 & = \frac{1 - (2\xi/L_h-1)^2}{\beta_1^2\sin{\eta}^2 + \beta_2^2\cos{\eta}^2}\beta_1^2\beta_2^2\,,\\
  \beta_1 &= r_1 - \hat{r}\sin{\left(4\xi/L_h - 2 \right)}\,, \\
  \beta_2 &= r_2 + \hat{r}\sin{\left(4\xi/L_h - 2 \right)}\,.
\end{align}
We specify the constant morphological parameters as $r_1=0.5\si{\um}$,
$r_2=1.25\si{\um}$, and $\hat{r}=0.28\si{\um}$, representing a model human
spermatozoon following \citet{Ishimoto2014}. \changed{Such parameters yield
ellipsoidal cross sections with maximal major and minor axes of
approximately 2.68\si{\um} and 1.28\si{\um} respectively, as shown in
\cref{fig:methods:mesh:head}.}


\end{document}